# Extraordinary magnetization behavior of single crystalline TbFe$_{4.4}$Al$_{7.6}$

N.P. Duong[*], E. Brück, P.E. Brommer, A. de Visser, F.R. de Boer and K.H.J. Buschow

Van der Waals-Zeeman Institute, University of Amsterdam,

Valckenierstraat 65, 1018 XE Amsterdam, The Netherlands

**Abstract**

We report the observation of a field-induced transformation from the easy-plane antiferromagnetic structure to the easy-axis ferrimagnetic structure in a single crystal of TbFe$_{4.4}$Al$_{7.6}$ (tetragonal ThMn$_{12}$ structure) at 5 K. Such a field-induced, irreversible transition has been identified for the first time. This transition is accompanied by a giant orthorhombic distortion: $e_{aa} = - e_{bb} \approx 3.5 \times 10^{-4}$ that is associated with a magnetic hardness ($\mu_0 H_C \approx 3$ T) that is unprecedented in this category of materials.

[*] Corresponding author. Tel: +31-20-525-5794; fax: +31-20-525-5788

*E-mail address*: pdnguyen@science.uva.nl (N.P. Duong).

## 1. Introduction

In a previous investigation, we have studied the magnetic properties of a single crystal of TbFe$_{4.4}$Al$_{7.6}$ [1]. This compound belongs to the large class of ternary rare earth compounds with the tetragonal ThMn$_{12}$ structure [2]. When measured with the field applied along the *c* direction, the magnetization is comparatively low. It slightly increases below about 50 K, and there is a shallow maximum around 160 K. From results of neutron diffraction and Mössbauer spectroscopy performed on similar ThMn$_{12}$-type compounds, it can be inferred that these temperatures mark the onset of magnetic ordering of the Fe sublattice and rare-earth sublattice, respectively [2-5]. When the field is applied along the [100] direction in the basal plane, the magnetization is much larger and its temperature dependence is characterized by pronounced thermal-history effects. Measurements of the magnetic isotherms at various temperatures show these effects to be due to the occurrence of coercivities that are extremely large at low temperatures but decrease at higher temperatures. These coercivities develop after a field-induced transition, the easy direction of the net magnetization being along the [100] direction.

Surprisingly, a strong hysteresis is observed perpendicular to the uniaxial direction of the underlying tetragonal crystal structure [1]. It is well known from many studies of various types of rare-earth-based permanent magnets that magnetization reversal in negative fields is prevented by pinning of Bloch walls or by features preventing the nucleation of reversed domains. The prerequisite for this behaviour is a crystal structure with uniaxial symmetry (hexagonal or tetragonal) and a strong magnetocrystalline anisotropy that aligns the magnetization in the uniaxial direction. Materials that have their easy magnetization direction perpendicular to the uniaxial direction are known not to sustain any coercivity. In tetragonal materials, this occurs because of the equivalence of the *a* or [100] direction and the *b* or [010] direction: one may imagine that the wall energy for Bloch walls (parallel to the easy plane) is rather low.

In the present investigation we have studied the magnetic properties of the same TbFe$_{4.4}$Al$_{7.6}$ single crystal in more detail in order to trace the origin of this unusual behavior. For the magnetic properties and the preparatory conditions we refer to the earlier publication [1].

The anisotropic properties were studied in magnetic fields up to 9 T, applied along various directions of the single crystal, in an Oxford Instruments Maglab magnetometer. The field dependence of the magnetization at 5 K measured along the *c* direction ([001]) is shown in Fig. 1. The magnetization remains low and shows no hysteresis. Under the assumption that the Tb moments are more or less confined to the easy plane, this magnetization is mostly due to the Fe moments, under the influence of the applied field and the molecular field exerted by the Tb moments. A completely different behavior emerges, however, when the measurements are performed in a direction perpendicular to the *c* axis. It can be seen in Fig. 1 that, at a critical field $\mu_0 H_{cr[100]} = 4.2$ T, the virgin magnetization curve displays a field-induced transition to a state with a large net magnetic moment. When decreasing the field strength after reaching 9 T, this magnetic moment is retained even for negative fields. Such an extremely strong magnetic hysteresis and the concomitant high coercivity are generally encountered in permanent-magnet materials. However, unlike in permanent magnets, the virgin magnetization curve falls mainly outside the hysteresis loop in the present material. A similar behavior, although much less pronounced, has been observed recently in a DyFe$_4$Al$_8$ single crystal [5].

We interpret this field-induced transition in the [100] isotherm in Fig. 1 as breaking of the antiferromagnetic order in both the Tb sublattice and the Fe one. This leads to a ferrimagnetic state with antiparallel Tb and Fe sublattice moments, a ferrimagnetic state that has been observed by powder neutron diffraction to be present already in zero field in the RFe$_5$Al$_7$ compounds which have slightly higher Fe concentration [6]. Once magnetized in the *a* direction with an applied field larger than the critical field, the single crystal behaves like a small permanent magnet with a remanent moment equal to $M_r([100])$ = 3.20 $\mu_B$/f.u.. This is more clearly seen in Fig. 2 where we show the remanence in zero field as a function of the measuring direction in the basal plane. Note that the magnetization is zero in the *b* direction, showing that the *a* ([100]) and *b* ([010]) directions are no longer equivalent in this originally tetragonal compound.

Revealing in this respect is also the magnetization behavior shown in Fig. 1, with the field applied in the [110] direction. The results shown in Fig. 1 are consistent with the following picture: in a field higher than, say, 7 T the material is forced into the ferrimagnetic state described above, partititioned in domains in which either the *a* direction or the (original) *b* direction is the easy direction probably in equal portions. Moreover, if we assume that the in-plane anisotropy is very large (in line with the permanent-magnet like behavior mentioned above), the (zero field) remanent moment observed in the [110] direction should be about $M_{r[100]}/\sqrt{2}$ = 2.26 $\mu_B$/f.u., in fair agreement with the results shown in Fig. 1.

In this connection, it is interesting to investigate the field dependence of the magnetization with the field applied in the *b* direction, after the crystal has been fully magnetized in the *a* direction. Preliminary results are shown in the inset of Fig. 1. Initially, the magnetization increases only very modestly with increasing field strength, suggesting an in-plane magnetocrystalline anisotropy of substantial magnitude. The slope is about twice that of the measurement in the *c* direction, so the in-plane anisotropy is much smaller than that for the *c* direction. In fact, now the net magnetic moment along the *b* direction probably consists of a positive Tb contribution opposite to the Fe contribution. For a value of the field along *b* somewhere between 3 T and 4 T, a jump-like increase of the magnetization occurs, up to values similar to those obtained in the *a* direction when originally magnetizing the crystal in the latter direction. It is also seen that a remanence of 3.2 $\mu_B$/f.u. has been attained, now along *b*. This means that the relative role played by the *a* (*b*) directions can be changed by applying a field larger than about 3 T along the *b* (*a*) directions.

One can summarize the results described above as follows: Before magnetizing the crystal, both in-plane <100> directions are equivalent. But after magnetizing the crystal in a field larger than $B_{cr[100]}$ along one of these directions, this direction has become a magnetic easy axis. This suggests that the field-induced transition not only breaks the antiferromagnetism of the R- and Fe sublattices, but simultaneously breaks the symmetry of the crystal lattice. Magnetostriction measurements to be reported below show that this is the case, indeed.

The magnetostriction was measured with applied fields along the *a* [100] direction. The top panel of Fig. 3 shows the relative length change $\Delta l/l$ of the single crystal measured along the *a* direction, and the bottom panel shows the results along the *b* [010] direction. The bottom part of Fig. 3 is almost a mirror image of the top part. This means that the relation $e_{aa} = - e_{bb}$ holds during the whole magnetization process. Notice that other deformation modes (such as *e.g.* $e_{cc}$, in the present case equal to the induced volume magnetostriction by virtue of the experimental result $e_{aa} + e_{bb} = 0$) cannot be determined in this set-up. The field-induced transition leads to a pronounced increase in length, up to $e_{aa} - e_{bb} \approx 700$ ppm, of the same order of magnitude as found in some (Tb,Fe) compounds, although still not quite reaching the values found in *e.g.* terfenol, a giant-magnetostrictive material based on the compound $TbFe_2$. Comparing with the loop behavior shown in Fig. 1, one sees that the hysteresis loop in the magnetostriction measurement is considerably broader than that in the magnetization measurements. Evidently, the magnetization processes in these two measurements are different. Experimentally, in the magnetization measurement configuration, it is well possible to orient the single-crystal sample precisely, and to check this by X-ray Laue pictures. In the magnetostriction cell, however, it is difficult to ascertain the orientation better than within about 0.1 rad. It is remarkable that such a tiny misorientation would have such a big influence. We return to this problem below. Nevertheless, it is apparent that the increase in length is lost during the magnetization reversal, but is fully recovered after completion of this process. The results show that the lattice is transformed from tetragonal to orthorhombic, and that the deformation is irreversible after a threshold (or transition) is passed. The elongated direction becomes the in-plane easy direction. The *c* direction, of course, remains 'the' hard direction.

In principle, magnetostriction data, as shown in Fig. 3, offer an excellent opportunity to obtain detailed information on the magnetization reversal mechanism. Bearing in mind the differences in experimental accuracy mentioned above, we will now compare the magnetization loop obtained for the field in the [100] direction (solid line in Fig. 1) and the magnetostriction loop shown in Fig. 3a, *i.e.* $e_{aa}$ - $e_{bb}$ (see above), presumably to be correlated with the average over all domains $<M_a^2 - M_b^2>$. Initially, the 'virgin' magnetostriction curve may well show this quadratic dependence on the magnetization. The fact that the "jump" doesn't show up in the magnetostriction and that saturation is reached at about 7 T can presumably be ascribed to the misorientation. Magnetization reversal sets in for negative fields of about 2 T. Fig. 3a shows that this is accompanied by a steadily growing decrease in $\Delta l$. Since magnetic domains of the type [100↑] (*M* along the positive *a* axis) and their reversed counterparts [100↓] (*M* along the negative *a* axis) have both the same positive $\Delta l$ value (see the extreme right and left parts of Fig. 3a) this means that - at least for the misoriented magnetostriction sample - magnetization reversal does not proceed by means of [100↓] domains growing at the expense of [100↑] domains. The steadily growing decrease in $\Delta l$ therefore requires the presence of an increasing volume of *b*-domains, *i.e.* domains of the type [010→] and/or [010←], at the expense of [100↑] domains. The contribution of the *b*-domains to the magnetization measured along *a* is small, but not negligible, as can be read off from the inset in Fig. 1. Since the magnetostriction presumably is proportional to ($M_{Tb,[100]}^2 - M_{Tb,[010]}^2$), the contribution of the *b*-domains is almost opposite to that of the *a*-domains. In fact, the point where $\Delta l$ passes through zero in Fig. 3a can be characterized as consisting of almost equal amounts of *a*- and *b*-domains. When Δ*l* becomes negative, the *b*-domain volume does exceed the *a*-domain volume. Δ*l* reaches only 25% of its maximum negative value, so there must be *a*-domains in this situation. Without further modelling, it is hard to say whether reversal to [100↓] domains has already taken place. Nevertheless, when the magnetostriction goes on to become more positive again, these [100↓] domains must be formed, in competition with the *b*-domains. From the inset of Fig. 1 it can be derived that the fields required for this process are larger than 3 T, in agreement with the magnetostriction data.

Finally we wish to discuss briefly a possible origin of the magnetoelastic phenomena observed in the present investigation. It is well known that magnetically ordered rare-earth compounds can give rise to strong magnetoelastic phenomena. A well known example is terfenol, a giant-magnetostrictive material based on the compound $TbFe_2$. As discussed in much detail by Morin and Schmitt [7], a substantial part of the magnetostriction in rare-earth-based materials is due to the direct coupling between the deformation of the lattice and the aspherical 4f-charge cloud. This can conveniently be described by a single-ion magnetoelastic Hamiltonian which can be considered as the strain derivative of the crystal-field Hamiltonian for the present symmetry. The main term in the latter Hamiltonian takes the form [5]

$$H^\lambda = - \sqrt{3} \bullet B^\lambda e_2^\lambda O_2^2,$$

where $B^\lambda$ is the magnetoelastic coefficient and where the appropriate strain component in the basal plane can be expressed as $e_2^\lambda = 1/2\sqrt{2} \bullet (e_{aa} - e_{bb})$. The temperature dependence of the magnetoelastic effects is determined by the thermal average $<O_2^2>$ of the Stevens operator [7]. It can be shown that $<O_2^2>$ falls off with temperature as $m_{Tb}^3$, where $m_{Tb}$ is the reduced terbium-sublattice moment $M_{Tb}(T)/M_{Tb}(0)$ [8]. We discussed above that the magnetic hardness, as embodied by the coercivity $H_C$, is intimately related to the occurrence of the orthorhombic distortion. Unfortunately, no experimental data are available for $M_{Tb}(T)$, but the presently determined large magnetoelastic effects and their expected strong temperature dependence offer a convenient explanation for the rapid decrease of $H_C$ with temperature reported previously [1].

**Figure captions**

Fig. 1. Field dependence of the magnetization at 5 K measured on a single crystal of the approximate formula composition TbFe$_{4.4}$Al$_{7.6}$ with the field applied in the [100], [110] and [001] direction. The inset shows the field dependence in the [010] direction after the crystal had been magnetized in a field of 9 T in the [100] direction and the field has subsequently been reduced to zero.

Fig. 2. Dependence of the remanence at 5 K measured in zero field in various directions in the basal plane on a single crystal of the approximate formula composition TbFe$_{4.4}$Al$_{7.6}$ after the crystal had been magnetized in a field of 9 T in the [100] direction and the field has subsequently been reduced to zero.

Fig. 3. Field dependence of the magnetostriction at 5 K measured (a) in the [100] direction and (b) in the [010] direction on a single crystal of the approximate formula composition TbFe$_{4.4}$Al$_{7.6}$ with the field applied along the [100] direction.

Fig. 1

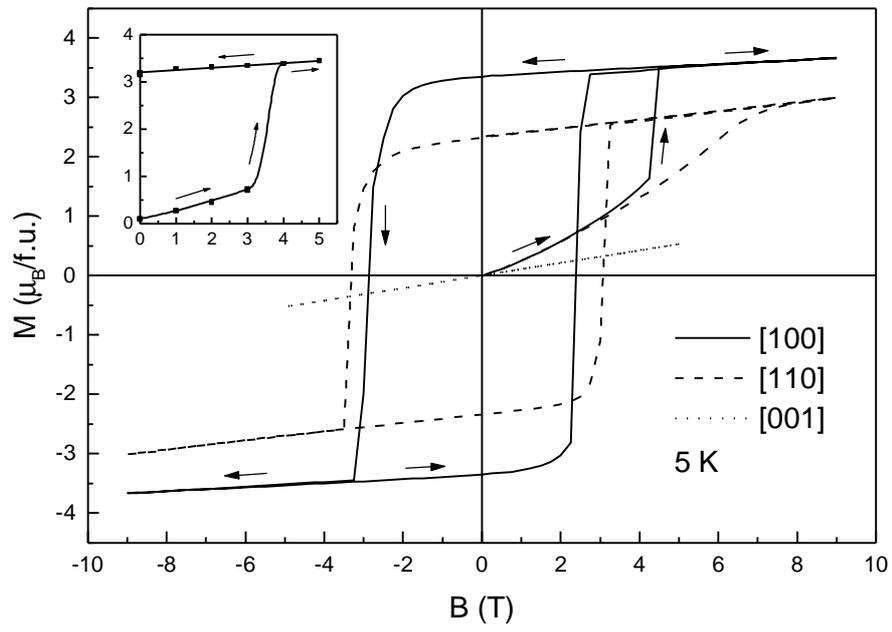

Fig. 2

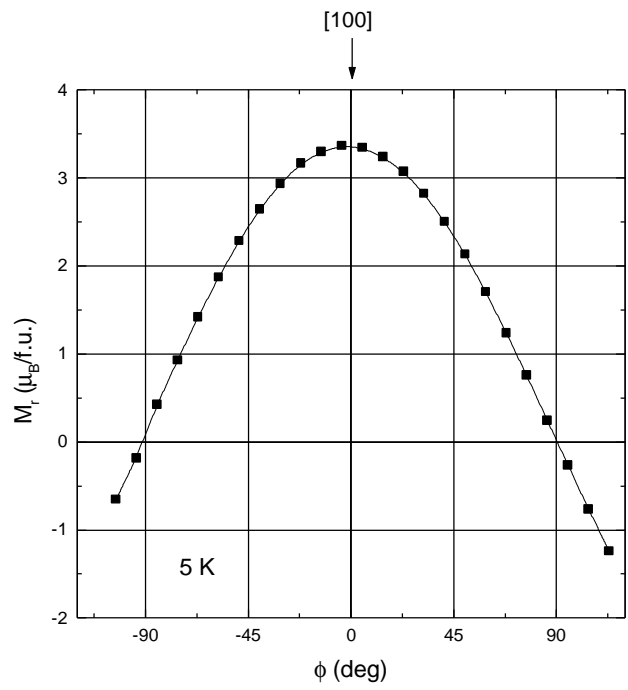

Fig. 3

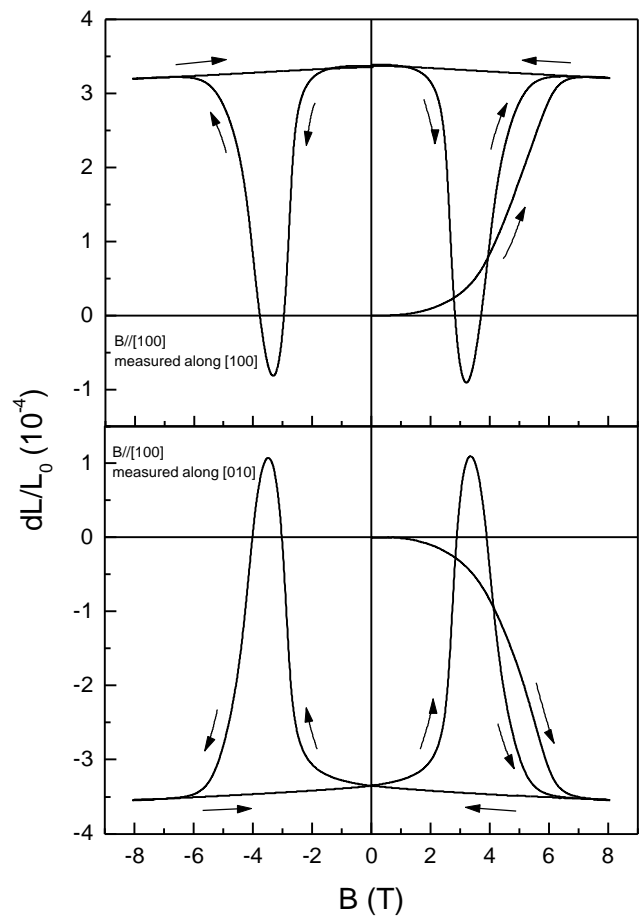